\documentclass[conference]{IEEEtran}
\IEEEoverridecommandlockouts
\usepackage{cite}
\usepackage{comment}
\usepackage{amsmath,amssymb,amsfonts}
\usepackage{algorithmic}
\usepackage{graphicx}
\usepackage{textcomp}
\usepackage{xcolor}
\def\BibTeX{{\rm B\kern-.05em{\sc i\kern-.025em b}\kern-.08em
    T\kern-.1667em\lower.7ex\hbox{E}\kern-.125emX}}

\usepackage{adjustbox}
 \usepackage{xltabular}
\usepackage{booktabs}
\usepackage{ragged2e}
\usepackage{caption}
\usepackage{url}

\usepackage{nicematrix}
\usepackage{pgfplots}
\usepackage{tikz}
\usepackage[dvipsnames]{xcolor}
\pgfplotsset{compat=1.18}
\usepackage[most]{tcolorbox}

\begin{document}

\title{Identifying Deceptive Patterns Across Three Age Groups: A Heuristic-Based Cognitive Walkthrough Study of Mobile Apps\\
}

\author{\IEEEauthorblockN{\textsuperscript{} Nasra Hassan}
\IEEEauthorblockA{\textit{}
\textit{Carleton University}\\
Ottawa, Canada \\
Nasrahassan3@cmail.carleton.ca}
\and
\IEEEauthorblockN{\textsuperscript{} Hala Assal}
\IEEEauthorblockA{\textit{}
\textit{Carleton University}\\
Ottawa, Canada \\
HalaAssal@cunet.carleton.ca}
}

\maketitle

\section{Abstract}
Deceptive patterns are tactics used to manipulate users into performing unintended actions. Today, many of these deceptive patterns are implemented in mobile apps targeting diverse age groups. In this paper, we employ a heuristic-based cognitive walkthrough to explore how deceptive patterns are tailored to three age groups, specifically teens (12-17), adults (18-49), and older adults (50+), across different app categories. By analyzing 30 apps spanning 6 categories, we found that 93\% of these apps use the \textit{nagging} pattern. Furthermore, our findings reveal that entertainment apps contain significantly more deceptive patterns than other app categories, such as music/books. Our data also shows that entertainment apps for older adults use \textit{sneaking} patterns more frequently than entertainment apps for teens or adults. These findings call for the development of more ethical, age-specific design guidelines to protect users from targeted digital manipulation attempts.

\section{Introduction}
Deceptive patterns (also referred to as dark patterns) have become a major concern for users of various ages. With today's modern technology, it has become easier for designers to implement these patterns through mobile apps more than ever. Many apps across various categories intentionally steer their users into performing actions that are not in their favour. Mathur et al. \cite{mathur2019taxonomy} investigated $\sim$11K shopping websites that apply deceptive patterns to nudge users into making additional purchases or simply prompt them to share more information. Similarly, Luguri and Strahilevitz \cite{luguri2021darkpatterns} demonstrated that deceptive patterns exploit customers' cognitive biases to manipulate their decision-making. This leaves users in a state of confusion or regret, specifically when encountering patterns such as \textit{hidden information}, \textit{trick questions}, and \textit{obstruction}. 

However, much of this previous work laid the foundations for studying deceptive patterns by creating taxonomies (such as Gray et al. \cite{gray2018dark}). Furthermore, Di Geronimo et al. \cite{di2020ui} implemented a different approach by heavily relying on analyzing popular mobile apps and recording examples of identified patterns broadly, finding that most apps contain one or more forms of deceptive patterns. Despite these efforts, there is a current gap in the literature regarding how apps vary their use of deceptive patterns across different target age groups, and whether the implementation of these patterns differs by app category. To address this, we conducted a heuristic-based cognitive walkthrough study \cite{privitera} analyzing 30 apps tailored to different age groups across 6 categories, including shopping, gaming, and health \& fitness. In our study, we address the following research question: \textbf{\textit{How do deceptive patterns vary across different app categories and target age groups?}} By understanding how such patterns are used across multiple apps and categories, our objective is to contribute to a more comprehensive understanding of deceptive pattern practices across different age groups.

\section{Background and Related Work}
Early work in deceptive patterns introduced by Brignull \cite{HarryBrignull} established the field's core concepts, with standardized taxonomies later developed \cite{gray2018dark,gray2024ontology}. With such foundational work, subsequent research has studied deceptive patterns from various lenses, such as examining how they are used in gaming apps \cite{zagal} or in specific social media platforms like Instagram \cite{instagram}, and the impacts of these patterns on users \cite{seyson2025exploring}. Users encounter deceptive patterns more frequently---often with limited understanding or recognition abilities; therefore, these tactics are widely considered malicious design practices. Below, we discuss distinct angles through which deceptive patterns have been studied.

\textbf{What are Deceptive Patterns?}
The term was originally introduced by cognitive scientist and designer Harry Brignull \cite{HarryBrignull}, who defined deceptive patterns as \textit{``a user interface that has been carefully crafted to trick users into doing things... they are not mistakes, they are carefully crafted with a solid understanding of human psychology, and they do not have the user’s interests in mind''} (as cited in \cite{gray2018dark}). This often comes in many forms, such as pressuring users into actions they are not initially intending to execute. Examples include forcing users to register for an account just for browsing purposes, or making it intentionally difficult to find relevant information. 

\section{Taxonomies and Approaches to Deceptive Patterns}
To address the rising concerns of deceptive patterns, many researchers have defined and systematically classified these patterns. For example, deceptive patterns have been categorized into various sections, such as obstructing information or socially engineering users, emotionally impacting them, and causing a sense of urgency. 

Gray et al. \cite{gray2018dark}, for example, identified five primary categories of deceptive patterns to create a foundational taxonomy. These include \textit{obstruction, sneaking, nagging, interface interference, and forced action}. Building on this, Gray et al. \cite{gray2024ontology} later classified deceptive patterns present in modern technology into low-, meso-, and high-level patterns. This classification provides a more structured, in-depth understanding (detailed in Section~\ref{sec:HML}), going beyond designers' general intent by highlighting concrete tactics. In addition, Mathur et al. \cite{mathur2021makes} introduced a new attribute called \textit{disparate treatment}, defined as disadvantaging or treating one group of users differently from another. This attribute further motivates our investigation into how the implementation of these deceptive patterns impacts distinct age groups.

\subsection{Deceptive Pattern Levels}
\label{sec:HML}
To evaluate how deceptive patterns impact different age groups, this study used Gray et al.'s \cite{gray2024ontology} three levels of hierarchy (low-, meso-, and high-level patterns). Employing these hierarchical levels helped with analyzing any present deceptive patterns and identifying the vulnerabilities of specific age groups.

\paragraph{Low-level patterns} These form visual and temporally placed user-interface tactics that a user can actually see and interact with \cite{gray2024ontology}. Examples include \textit{sneak into basket} and \textit{disguised ads}, where ads are incorporated as part of the app's features to look like a button or regular website content. For instance, a user may unintentionally interact with a promotional banner that mimics a real download button, thereby requiring extra effort to return to their page. Analyzing deceptive patterns under this level helps to examine these visual patterns and the limitations for specific age groups, like older adults.  

\paragraph{Meso-level patterns} These explain how the high-level patterns are carried out \cite{gray2024ontology}. For example, an implementation of the \textit{forced action} high-level pattern can be \textit{forced registration}, where users are pushed to create accounts as a way to compel them to share personal information, which may not be necessary to receive the service. Investigating the meso-level for age group comparisons adds more value, enabling an analysis of how different age groups' expectations are not met when navigating apps. 

\paragraph{High-level patterns} These explain the overarching strategies deployed to deceive users' intentions and manipulate their decision-making. While ``low-level" patterns show manipulation visually through interface elements, ``high-level" patterns highlight the designer's overall intent to trick users, making it difficult to identify the underlying deception. For age-related analysis, these ``high-level" patterns provide insights into the general strategies that include manipulative, coercive, and deceptive elements. Using these patterns can help in understanding user limitations and decision-making between different age groups. To contextualize these ``high-level'' patterns, Gray et al.'s updated ontology \cite{gray2024ontology} refined the five primary deceptive patterns originally introduced by Gray et al. \cite{gray2018dark}. Thus, this updated framework classifies these five foundational categories as ``high-level" patterns as part of a broader taxonomy. These categories include: 

\textit{\textbf{Sneaking}} patterns intentionally manipulate users by delaying, hiding, or disguising relevant information that, if available, would alter the user's decision \cite{gray2024ontology}. 
Patterns such as \textit{drip pricing, hidden costs, or partitioned pricing} (which hide the true costs of an item/service until the final step) are included here, manipulating users into transactions that would potentially be declined if full transparency were provided. 

\textit{\textbf{Obstruction}} patterns cause difficulties in task flow, making it more challenging to execute tasks with the intention of dissuading a user from taking an action \cite{gray2024ontology}. Examples include the \textit{roach motel} or \textit{immortal accounts} patterns, which deceptively provide a seamless subscription registration, and requiring extra measures to opting out. This causes redundant complexity for users, particularly once they have shared their data. 

\textit{\textbf{Interface Interference}} patterns favour specific actions for users to take over other options, therefore causing confusion and limiting discoverability \cite{gray2024ontology}. This includes implementing patterns like \textit{visual prominence}, where options such as accepting an offer are presented in a big, bright button, in contrast to the opt-out option, which may be shown in small, plain text. 

\textit{\textbf{Forced Action}} patterns pressure users to knowingly or unknowingly execute specific tasks, which oftentimes require additional steps, steering them away from their intended interaction \cite{gray2024ontology}. \textit{Forced registration} is one instance of this high-level pattern, restricting users from exploring an app without inputting any personal information. This essentially converts a voluntary service into a mandatory data or financial transaction.

\textit{\textbf{Social Engineering}} patterns provide options or information that would steer a user to take an action based on their individual and/or social cognitive biases, as a result, causing users to follow imposed or expected social norms \cite{gray2024ontology}. Patterns like \textit{high demand} (using emotionally deceptive language such as ``100 bought in 24h") take advantage of social proof, leading to a sense of urgency and social cognitive biases among users.

Ultimately, Gray et al.'s \cite{gray2024ontology} taxonomy creates a foundation for understanding deceptive patterns and identifying them. More importantly, it has built common knowledge not just among researchers but users as well, helping them understand how they may be manipulated. For example, the ability of users to recognize and understand \textit{nagging}, or to identify when key information is hidden from them during the execution of essential tasks---which would have resulted in a different action if not hidden---is vital for avoiding exploitation. 

\subsection{Digital Platforms and Deceptive Patterns} 
\label{sec:Digital Platforms}
Brignull’s \cite{HarryBrignull} original typology introduced early deceptive pattern concepts such as \textit{privacy zuckering} and \textit{hidden costs}; however, subsequent studies have analyzed diverse deceptive patterns present across commonly used apps. These empirical evaluations confirm that the categories covered by Gray et al. \cite{gray2024ontology} are identifiable for practical empirical analysis, used to provide a better understanding. 

For instance, Seyson and Willett \cite{seyson2025exploring} addressed the limitations in the literature regarding social networking services, specifically the lack of longitudinal studies analyzing deceptive patterns by investigating designs within Instagram’s \cite{instagram} user interface. From 2010 to 2024, researchers analyzed deceptive design strategies imposed on Instagram users, such as limiting users’ illusory control and delaying consent prompts, underscoring Instagram’s limitations in policy implementation \cite{seyson2025exploring}.

Similarly, Di Geronimo et al. \cite{di2020ui} found that despite 95\% of analyzed apps containing one or more deceptive patterns, users were generally unaware of these deceptive patterns---a concept referred to as ``DP-blindness.'' This lack of awareness highlights that users face substantial challenges in identifying such patterns in real-world interactions with various apps. Furthermore, deceptive patterns are not solely present on Western digital platforms. Hidaka et al. \cite{hidaka2023linguistic} performed a large study analyzing 200 Japanese Android apps and found 95\% of the apps embedded at least one deceptive pattern, averaging 3.9 per app. 

This average is substantially lower than what Di Geronimo et al. \cite{di2020ui} observed for English apps. The latter study underlined that, on average, 7.4 deceptive patterns were seen per app. These findings demonstrate that deceptive patterns appear systematically across various cultures and platforms, requiring a closer examination of the contextual elements that enhance their design and influence. 

\section{Deceptive Patterns in Cultural Variations Context}
Previous studies have demonstrated that deceptive patterns are widespread and not limited to Western apps. The way these deceptive strategies are executed, however, often correlates with cultural norms and platform purposes.

In Hidaka et al. \cite{hidaka2023linguistic}, researchers introduced a deceptive pattern distinctive from previous studies. \textit{``Linguistic Dead Ends''} is a deceptive pattern that was observed in Japanese Android apps. It introduced the idea that a design can limit users' ability to understand the content presented in the interface by imposing manipulative language. Since this pattern is deployed within a specific culture, it directly misleads users and prevents them from making an informed decision \cite{hidaka2023linguistic}.

Additionally, Hidaka et al. \cite{hidaka2023linguistic} and Di Geronimo et al.'s \cite{di2020ui} findings emphasize the importance of having transparent platform policies, cultural designs, and language choices, which significantly impact the way deceptive patterns are imposed and perceived. While Western-based deceptive pattern studies do not explicitly use the term \textit{``Linguistic Dead Ends''}, many underline similar foundations that are included in established taxonomies, which emphasize challenges such as the implementation of misleading or incorrect language. 

\section{User Awareness of Deceptive Patterns}
Research has demonstrated users' vulnerabilities to deceptive patterns, often due to cognitive limitations. For example, some interfaces explicitly do not display direct interactive cues that would guide a user into thinking about deceptive patterns. 

As previously highlighted in Section~\ref{sec:Digital Platforms}, the study by Di Geronimo et al. \cite{di2020ui} introduced the ``DP-blindness'' phenomenon. This highlights the users' limitation in their ability to strategically recognize deceptive patterns imposed during their interaction with various apps. Similarly, the study ``Growing Up With Dark Patterns: How Children Perceive Malicious User Interface Designs'' \cite{schafer2024growing} observed that complicated trick questions often caused difficulties in children's ability to evaluate the patterns correctly, particularly as there were no direct outcomes visible to them, despite their familiarity with user interfaces. 

Sch\"afer et al. \cite{schafer2024growing} reported key observations from deceptive cookie consent designs that contained \textit{trick questions}. When children were asked to analyze these designs, many reported using terms such as ``fishy'', ``weird'', or ``difficult'' \cite{schafer2024growing}. This underlines that users, specifically younger individuals, often fail to make the psychological connections to understand why certain designs are implemented in a specific way unless they are guided by the interface through transparency or educational options.  

\section{Methodology}
In this study, we identify deceptive patterns in mobile apps through a heuristic-based cognitive walkthrough (CW) approach. While traditional cognitive walkthroughs are deployed to evaluate whether an interface allows users to easily complete tasks within a given system \cite{privitera}, our research adapts this methodology to examine user vulnerabilities and the implementation of deceptive patterns. Prior researchers such as Habib et al. \cite{habib2022okay} have also combined heuristic evaluation and cognitive walkthrough methods, helping them evaluate interfaces for deceptive patterns. To carry out our study, we followed the CW methodology, guided by a set of heuristics based on the analysis framework described below. Apps analyzed targeted different age groups: teens (12-17), adults (18-49), and older adults (50+). In our study, we systematically investigated each selected app to examine deceptive patterns implemented, helping to analyze their prevalence across age groups and app categories. Overall, we analyzed 30 apps across six categories (i.e., 5 apps per category). These covered the different age groups: 8 apps targeting teens, 17 targeting adults, and 5 targeting older adults. During our search, we found that certain app categories did not apply to specific age groups (e.g., we did not find a social media app specifically targeting older adults). While older adults likely use apps categorized as for adults, we chose to follow the app's age classification for consistency. Additionally, our methodology focuses on analyzing apps tailored \emph{specifically} to our target age groups. 

\begin{figure}[htbp]
         \centering
         \includegraphics[width=\columnwidth]{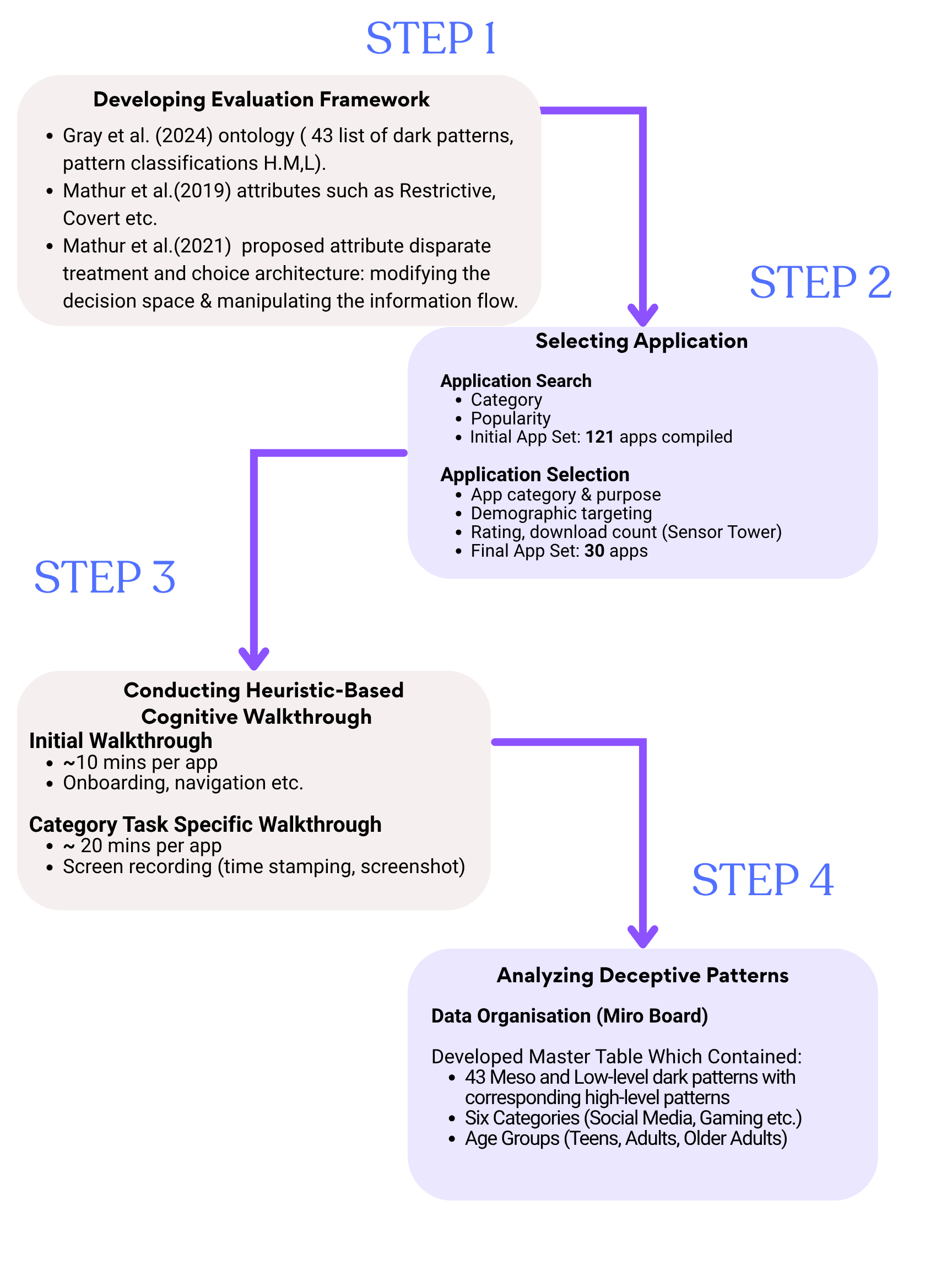}
         \caption{\textbf{Sequence of steps taken to execute heuristic-based cognitive walkthrough }}
         \label{fig:Flow Diagram} 
     \end{figure}

We first used existing taxonomies to develop an evaluation framework. Next, we identified and selected mobile apps for analysis, which was followed by the walkthrough to examine them for any deceptive patterns. Finally, we analyzed our findings to highlight any trends and understand the implementation of deceptive patterns among all age groups and app categories. We explain the specifics of our four-step methodology below (see Fig~\ref{fig:Flow Diagram}).

\paragraph{\textbf{STEP 1 Developing Evaluation Framework}}\label{sec:taxonymyCreation} 

Before analyzing any Google Play \cite{google_play} apps to identify repeating patterns and develop our evaluation guide, we relied on the framework proposed by Gray et al. \cite{gray2024ontology} as our primary source. Deceptive patterns in this study were divided into three levels of hierarchy (low-, meso-, and high-level patterns). This allowed us to use meso- and low-level patterns during our walkthrough to spot any deceptive patterns present. In addition, we initially reviewed Mathur et al.'s \cite{mathur2019taxonomy} taxonomy, which also highlighted the five attributes that explain how they modify a user's choice architecture. These include: asymmetric, covert, deceptive, information hiding, and restrictive. Although we initially included a section in our evaluation guide for the attributes each meso- and low-level pattern may fall under, we recognized during our walkthrough that this information was redundant as it frequently overlapped with Gray et al.'s \cite{gray2024ontology} work. Thus, we did not move forward using these attributes in our final analysis. 

To begin building this evaluation framework as a walkthrough guide during Step 1, we focused on including a few key details. First, we included the categorized deceptive patterns listed using the low-, meso-, and high-level ontology \cite{gray2024ontology}. We also initially included a section highlighting the deceptive pattern attributes and choice architecture \cite{mathur2019taxonomy}; however, as noted, we did not utilize these components during our walkthrough. Moreover, to prepare for the analysis stage in Step 3, we used the definitions Gray et al. \cite{gray2024ontology} established, pairing them with visual examples in the form of screenshots. These examples were extracted from various public websites, along with personal examples extracted from daily-used apps. While using these examples as a reference may have introduced a potential risk of confirmation bias, we solely used these screenshots for initial training and rapid recognition of deceptive patterns. We understood that deceptive patterns varied in their implementation depending on the interface (app vs. website), which is why we did not use these examples as a primary source. The overall purpose of this preliminary step was to provide contextual real-world scenarios without restricting our process.   

Furthermore, during this step, specific deceptive patterns from Gray et al.'s \cite{gray2024ontology} ontology were scoped to ensure we stay focused on deceptive patterns from a mobile app lens. For example, according to Gray et al. \cite{gray2024ontology}, \textit{Cuteness} is defined as using emotional or sensory manipulation to incorporate attractive cues in the design of a ``robot interface or form factor.'' We excluded the robots aspect as it falls outside our mobile app scope. Instead, we focused on how apps use the pattern of cuteness through built-in fictitious characters as a form of deceptive pattern to misguide users. These deployed characters are not necessarily deceptive themselves; however, they become a problematic deceptive design as apps utilize them to gain users' trust, manipulating user emotions into taking privacy-risky actions, for example. 

\paragraph{\textbf{STEP 2 Selecting Application}} A list of 121 apps from the Google Play store \cite{google_play} was gathered. These covered various categories (e.g., social media, shopping) and age group ratings (e.g., T for teens). While putting this list together, we aimed to include widely known and popular apps, as well as less popular ones. To ensure our heuristic-based cognitive walkthrough was manageable, we narrowed down the initial set of apps based on specific criteria such as category, popularity, and age target. Thus, our final selection contained 30 apps covering six categories: social media, shopping, entertainment, gaming, music/book, and health \& fitness. The selected apps covered different age groups and app popularity (measured by the number of worldwide downloads from Sensor Tower \cite{sensortower_app}) to avoid bias and ensure a diverse sample. 

After selecting our 30 apps, we began confirming the targeted age groups—teens (12-17), adults (18-49), and older adults (50+)—by examining app descriptions, articles, blog posts, and reviews. While some apps are used by all age groups, we carefully assessed each app to confirm under which age group it should be categorized and why. For example, with SHEIN, we anticipated this app would target adults, but after an initial walkthrough and confirming the rating (which was labelled as ``T" for teens), we proceeded with our analysis of this app for the teen group. Following this process throughout, we then documented the occurrence of the low- and meso-level deceptive patterns within each age group.

\paragraph{\textbf{STEP 3 Conducting Heuristic-Based Cognitive Walkthrough}}
\label{sec:DataRecording}

One researcher conducted the heuristic-based cognitive walkthrough for all apps. Each app was analyzed in a separate session lasting 30 minutes, producing a total of 15 hours of audio- and screen-recorded sessions. The researcher also took screenshots of identified deceptive patterns. Each walkthrough session was divided into two stages. First, an initial walkthrough for roughly 10 minutes was completed. This included browsing the main app features, creating an account, reviewing settings options, verifying exit behaviours by attempting to log out or close the app mid-task, interacting with personalization prompts, and getting a general overview of the app. 

The second stage of Step 3, lasting around 20 minutes, was a task-oriented walkthrough. These tasks were predetermined based on the app category (e.g., streaming video or audio content for entertainment apps). Once the walkthrough was completed, the researcher reviewed the recording and noted the observed deceptive patterns, including any that were missed. They also recorded the timestamp of the specific pattern identified to help with later reviews. 

\paragraph{{\textbf{STEP 4 Analyzing Deceptive Patterns}}}

Upon completion of Step 3, we evaluated our findings. To organize the  qualitative data, we compiled everything into a Miro Board \cite{miro}, using this to visually review all observed patterns and identify any trends. Within Miro \cite{miro}, we built a master table to cross-reference the 43 low- and meso-level deceptive patterns \cite{gray2024ontology} with the six app categories and three age groups. Using our framework from Step 1 helped us to study the frequency with which the low- and meso-level deceptive patterns appeared across each age group: teens (12-17), adults (18-49), and older adults (50+). 

Following data organization, the research team met to collaboratively review the identified deceptive patterns to ensure alignment. This step was vital to verify that the findings were properly categorized based on our criteria. However, we identified disagreements regarding the patterns' classification. We resolved these discrepancies by reviewing pattern definitions and taking part in ongoing collaborative discussions, allowing us to reach full consensus. Overall, the analysis completed in this step allowed us to systematically explore the occurrence of each deceptive pattern and its prevalence in each app category.

\subsection{Limitations and Interpretations}
While we minimized methodological limitations, we identified some challenges carrying out this study.

First, identifying the \textit{dead ends} pattern was challenging. We observed instances where it was unclear if app elements or links were unresponsive due to technical issues or if this was a form of \textit{dead ends}. We chose to include these examples because broken links restrict users' access and impact their experience the same way, regardless of the intent. However, we recognize these findings may have included instances of technical issues misidentified as user manipulation.

Second, some deceptive patterns, such as \textit{Trick Questions} and \textit{Sneak into Basket}, were not identified in our analysis. We expect that the amount of time we spent analyzing each app may not have provided the full extent of deceptive pattern implementation for the analyzed apps. Our analysis also does not necessarily reflect a typical user's day-to-day interactions or the experience of frequent users. However, our 15 hours of interactions with apps (spending 30 minutes per app) ensured that covering a good breadth of apps was feasible.

Lastly, our dataset lacks normalization across the three age groups. The distribution of deceptive patterns varied as our app selection strategy focused on selecting 5 apps per category initially, before categorizing their target age group. This categorization required external research and demonstrated that mobile apps heavily market to adults as the highest paying consumers \cite{subscription_gap}. As a result, our findings naturally contain more deceptive patterns for adults due to the large volume of apps in our dataset that impact them. 

\section{Results}
\label{sec:Findings}

\label{sec:Frequent low-level}
Overall, in our analysis, we revealed that deceptive patterns were found at least once in all 30 apps analyzed. Across the 5 high-level deceptive patterns, the two most \textbf{common} were \textit{forced action} and \textit{interface interference}. We visualized our findings through a heatmap {(Figure \ref{fig:all age groups})} to address the trends across the six app categories.

To ensure we can fairly compare our app categories, which vary in size, the data was first normalized. We recorded the presence of low- and meso-level deceptive patterns per app. Each deceptive pattern found for an app was recorded once; multiple instances were not accounted for to avoid skewing frequency. We then normalized our data by dividing each instance by the total number of low- and meso-level deceptive patterns that each high-level deceptive pattern contained. We then multiplied it by 5 (this is the total number of apps under each app category). In {(Figure \ref{fig:all age groups})}, we represent our findings of the 5 high-level deceptive patterns and their occurrence for each app category. Higher values (shown in darker colours) demonstrate that an app category implements various patterns within that high-level deceptive pattern.

\begin{figure}[htbp]
         \centering
         \includegraphics[width=0.4\textwidth]{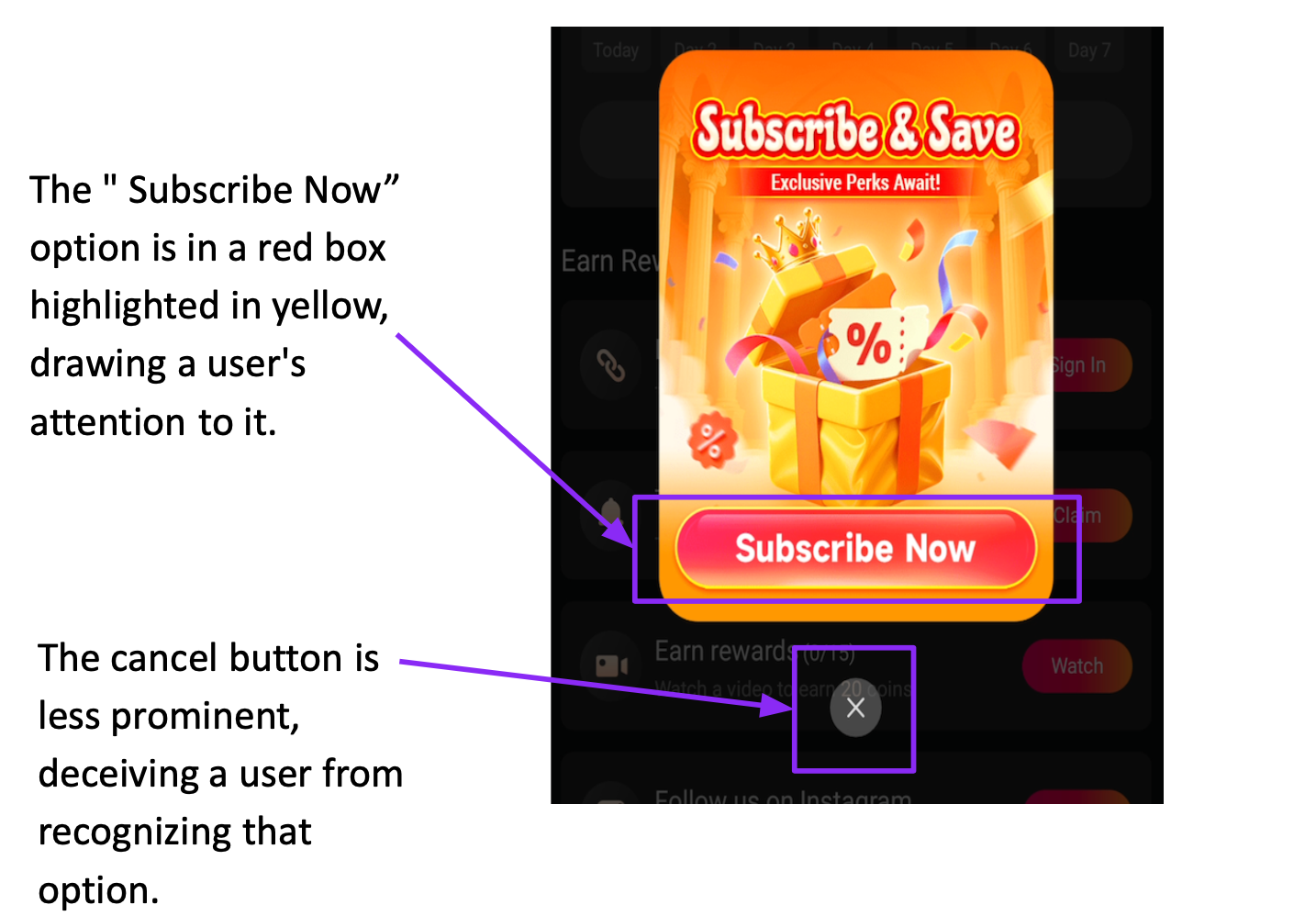}
         \caption{\textbf{DramaBox \cite{dramabox} continuously displaying a pop-up as a form of \textit{nagging} to force users to subscribe}}
         \label{fig:DramaBox pop-up} 
     \end{figure}

\begin{figure}[htbp]
         \centering
         \includegraphics[width=0.4\textwidth]{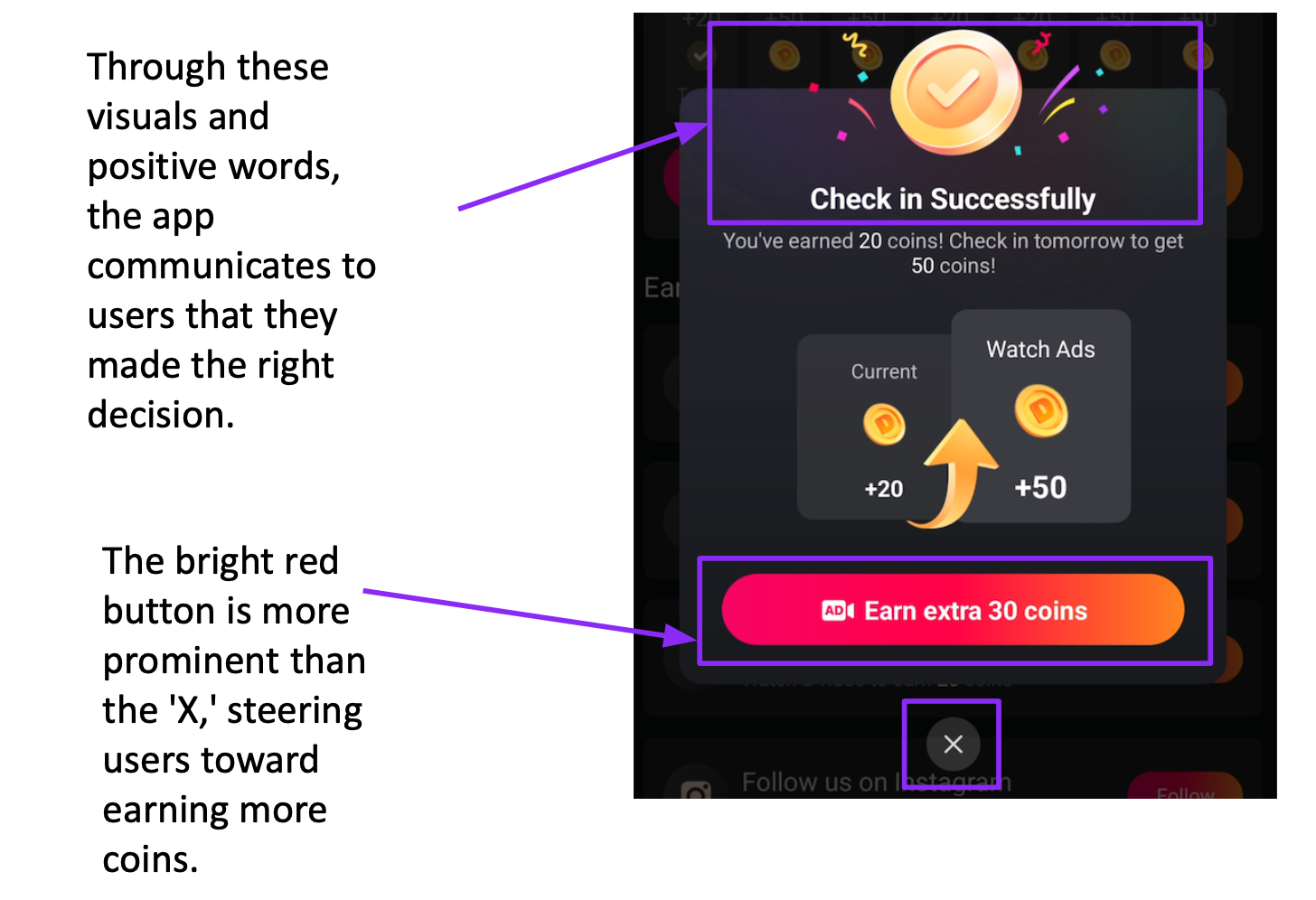}
         \caption{\textbf{DramaBox \cite{dramabox} using visual elements to get user's attention after watching an ad}}
         \label{fig:DramaBox VP} 
     \end{figure}

In detail, \textit{forced action} was implemented frequently through \textit{nagging}---observed in 28 of the 30 apps through persistent notifications and pop-ups (e.g., see Figure \ref{fig:DramaBox pop-up}). Additionally, \textit{Interface interference} was implemented through \textit{positive or negative framing, bad defaults}, and \textit{visual prominence}
(see Figure \ref{fig:clustered graph}). Across the three age groups, visual prominence (found in 23 apps) was deployed through animations and prominent colours nudging users towards interacting with specific elements such as offers and ads (e.g., see Figure \ref{fig:DramaBox VP}). \textit{Personalization, forced registration and bad defaults} deceptive patterns were equally prominent, as shown in (Figure \ref{fig:meso-low-dp}), each observed in 9 out of the 30 apps. 

\definecolor{heatRed}{HTML}{F27272}    
\definecolor{heatOrange}{HTML}{FFB366} 
\definecolor{heatYellow}{HTML}{FFEB9C} 

\begin{figure}[ht]
\centering
\scriptsize
\setlength{\tabcolsep}{2.5pt}
\begin{NiceTabular}{lcccccc}[cell-space-limits = 3pt]
Forced Action & \cellcolor{heatRed} 0.28 & \cellcolor{heatOrange} 0.14 & \cellcolor{heatOrange} 0.14 & \cellcolor{heatOrange} 0.12 & \cellcolor{heatOrange} 0.16 & \cellcolor{heatRed} 0.22 \\
Interface Interference & \cellcolor{heatRed} 0.22 & \cellcolor{heatOrange} 0.14 & \cellcolor{heatOrange} 0.14 & \cellcolor{heatOrange} 0.11 & \cellcolor{heatYellow} 0.06 & \cellcolor{heatOrange} 0.14 \\
Obstruction & \cellcolor{heatOrange} 0.16 & \cellcolor{heatOrange} 0.16 & \cellcolor{heatYellow} 0.04 & \cellcolor{heatYellow} 0.04 & \cellcolor{heatYellow} 0.08 & \cellcolor{heatYellow} 0.04 \\
Sneaking & \cellcolor{heatYellow} 0.10 & \cellcolor{heatYellow} 0.03 & \cellcolor{heatYellow} 0.07 & \cellcolor{heatYellow} 0.03 & \cellcolor{heatOrange} 0.17 & \cellcolor{heatYellow} 0.03 \\
Social Engineering & \cellcolor{heatYellow} 0.09 & \cellcolor{heatYellow} 0.07 & \cellcolor{heatYellow} 0.07 &  & \cellcolor{heatRed} 0.33 & \cellcolor{heatYellow} 0.04 \\

\RowStyle{\bfseries}
& Entertain. & Gaming & \shortstack{Health \&\\Fitness} & \shortstack{Music/\\Books} & Shopping & \shortstack{Social\\Media} \\
\end{NiceTabular}

\caption{\textbf{Occurrence of high-level deceptive pattern across all age groups for six app categories}}
\label{fig:all age groups}
\end{figure}

\begin{figure}[ht]
\centering
\scriptsize
\setlength{\tabcolsep}{2pt}
\begin{NiceTabular}{lcccccc}[cell-space-limits = 3pt]
Forced Action & \cellcolor{heatOrange} 0.14 & \cellcolor{heatYellow} 0.02 &   & \cellcolor{heatYellow} 0.04 & \cellcolor{heatYellow} 0.04 & \cellcolor{heatOrange} 0.14 \\
Interface Interference & \cellcolor{heatOrange} 0.11 & \cellcolor{heatYellow} 0.03 &    & \cellcolor{heatYellow} 0.02 & \cellcolor{heatYellow} 0.02 & \cellcolor{heatYellow} 0.08 \\
Obstruction & \cellcolor{heatYellow} 0.08 & \cellcolor{heatYellow} 0.04 &   &   &   & \cellcolor{heatYellow} 0.04 \\
Sneaking &   &   &   &   & \cellcolor{heatYellow} 0.03 &  \\
Social Engineering & \cellcolor{heatYellow} 0.04   & \cellcolor{heatYellow} 0.02 &   &   & \cellcolor{heatYellow} 0.09 & \cellcolor{heatYellow} 0.02 \\

\RowStyle{\bfseries}
& Entertain. & Gaming & \shortstack{Health \&\\Fitness} & \shortstack{Music/\\Books} & Shopping & \shortstack{Social\\Media}\\
\end{NiceTabular}
\caption{\textbf{Occurrence of high-level deceptive pattern for teens across six app categories}}
\label{fig:Teens HL}
\end{figure}

\begin{figure}[ht]
\centering
\scriptsize
\setlength{\tabcolsep}{2pt}
\begin{NiceTabular}{lcccccc}[cell-space-limits = 3pt]
Forced Action & \cellcolor{heatYellow} 0.08 & \cellcolor{heatYellow} 0.04 & \cellcolor{heatOrange} 0.12 & \cellcolor{heatYellow} 0.06 & \cellcolor{heatOrange} 0.12 & \cellcolor{heatYellow} 0.08 \\
Interface Interference & \cellcolor{heatYellow} 0.06 & \cellcolor{heatYellow} 0.05 & \cellcolor{heatOrange} 0.11 & \cellcolor{heatYellow} 0.09 & \cellcolor{heatYellow} 0.05 & \cellcolor{heatYellow} 0.06 \\
Obstruction & \cellcolor{heatYellow} 0.04 & \cellcolor{heatYellow} 0.08 & \cellcolor{heatYellow} 0.04 & \cellcolor{heatYellow} 0.04 & \cellcolor{heatYellow} 0.08 &  \\
Sneaking & \cellcolor{heatYellow} 0.03 &   & \cellcolor{heatYellow} 0.07 & \cellcolor{heatYellow} 0.03 & \cellcolor{heatOrange} 0.13 & \cellcolor{heatYellow} 0.03 \\
Social Engineering & \cellcolor{heatYellow} 0.04 & \cellcolor{heatYellow} 0.02 & \cellcolor{heatYellow} 0.04 &   & \cellcolor{heatRed} 0.24 & \cellcolor{heatYellow} 0.02 \\

\RowStyle{\bfseries}
& Entertain. & Gaming & \shortstack{Health \&\\Fitness} & \shortstack{Music/\\Books} & Shopping & \shortstack{Social\\Media}\\
\end{NiceTabular}
\caption{\textbf{Occurrence of high-level deceptive pattern for adults across six app categories}}
\label{fig:Adults HL}
\end{figure}

\begin{figure}[ht]
\centering
\scriptsize
\setlength{\tabcolsep}{2.3pt}
\begin{NiceTabular}{l | c | c | c | c | c | c}[
    cell-space-limits = 3pt,
    rules/color = gray!30, 
    rules/width = 0.2pt    
]
Forced Action & \cellcolor{heatYellow} 0.06 & \cellcolor{heatYellow} 0.08 & \cellcolor{heatYellow} 0.02 &   &   &  \\ \cline{1-1}
Interface Interference & \cellcolor{heatYellow} 0.05 & \cellcolor{heatYellow} 0.06 & \cellcolor{heatYellow} 0.03 &   &   &  \\ \cline{1-1}
Obstruction & \cellcolor{heatYellow} 0.04 & \cellcolor{heatYellow} 0.04 &   &   &   &  \\ \cline{1-1}
Sneaking & \cellcolor{heatYellow} 0.07 & \cellcolor{heatYellow} 0.03 &   &   &   &  \\ \cline{1-1}
Social Engineering &   & \cellcolor{heatYellow} 0.02 & \cellcolor{heatYellow} 0.02 &   &   &  \\ \hline

\RowStyle{\bfseries}
& Entertain. & Gaming & \shortstack{Health \&\\Fitness} & \shortstack{Music/\\Books} & Shopping & \shortstack{Social\\Media} \\
\end{NiceTabular}
\caption{\textbf{Occurrence of high-level deceptive pattern for older adults across six app categories}}
\label{fig:OA HL}
\end{figure}

\subsection{Deceptive Patterns in Unexpected App Contexts}

During our walkthrough, we observed unexpected deceptive patterns in some app categories. For example, the \textit{drip pricing, hidden costs, or partitioned pricing} deceptive pattern was identified in the social media category specifically, in Facebook \cite{Facebook}. The app's built-in ``Facebook Marketplace'' feature employed this manipulative pricing pattern by displaying items as ``Free'' and showing the true price only when selected. This tactic of delaying price details essentially limits transparency and can cause users to have inaccurate expectations.

\textit{Intermediate currency} in social media apps is another example of an unexpected deceptive pattern. Throughout our interaction with the Twitch app \cite{Twitch}, we observed a gifting option to support streamers through the use of virtual currency, which makes it hard to evaluate the true monetary costs of these transactions. We did not expect this, as virtual currency and monetization designs are more commonly discussed in gaming platforms \cite{zagal}. 

\subsection{Deceptive Patterns Across Age Groups}
In this section, we discuss how deceptive patterns target the different age groups. 

\begin{figure}[htbp]
         \centering
         \includegraphics[width=0.5\textwidth]{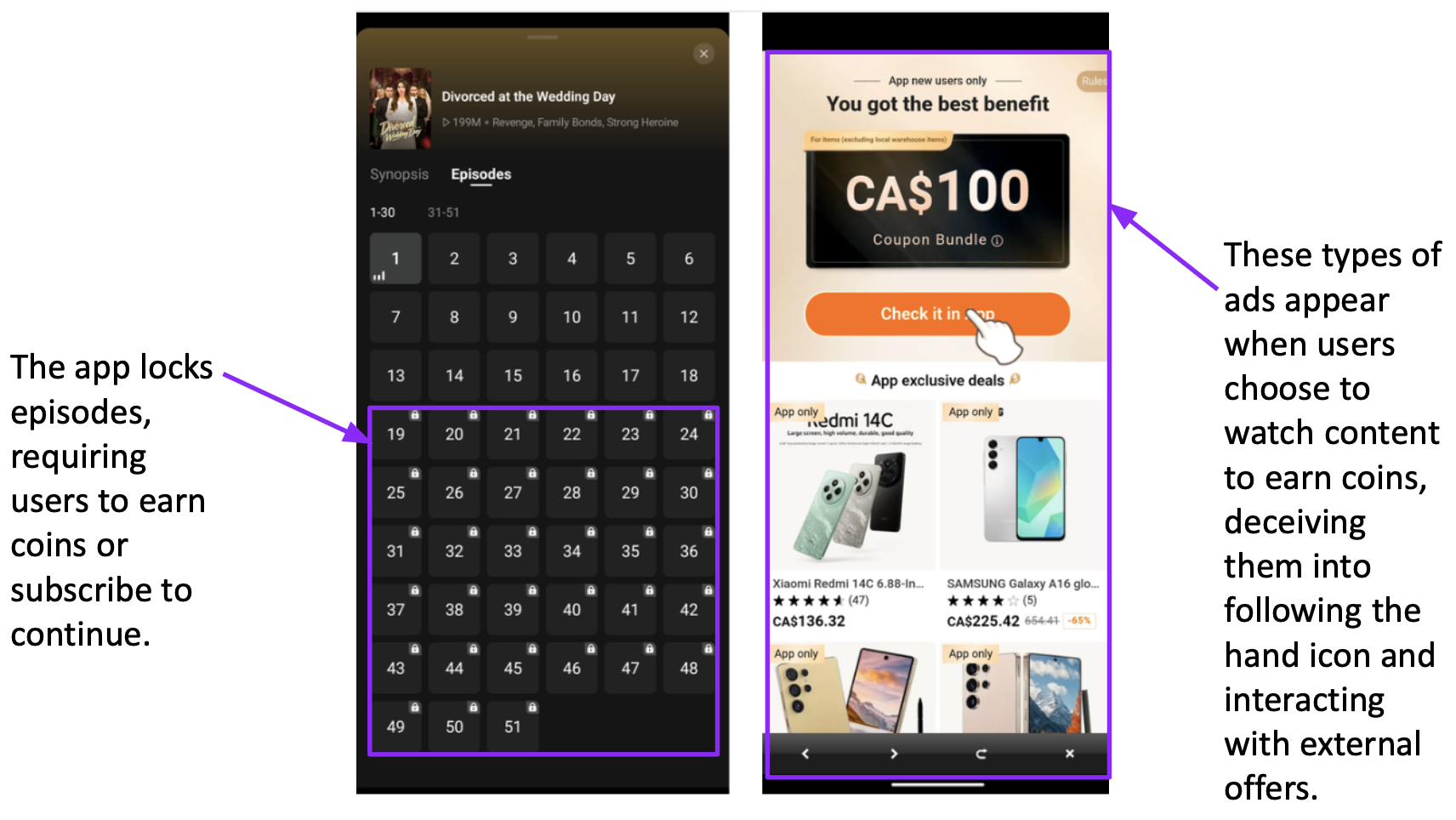}
         \caption{\textbf{DramaBox \cite{dramabox} app requiring users to watch ads if not subscribed to unlock episodes}}
         \label{fig:DramaBox Locked Episodes}
     \end{figure}

\paragraph{\textbf{Teens}}
Our analysis of the selected teen-oriented apps showed that this age group is exposed to \textit {forced action} patterns within both social media and entertainment apps. Following the same normalization procedure as Figure \ref{fig:all age groups}, Figure \ref{fig:Teens HL} highlights that social media and entertainment app categories deploy \textit{forced action} patterns the most, in contrast to other categories such as gaming (0.02). For \textit {forced action}, this is apparent through the use of patterns such as \textit{forced registration}, where users are frequently required to share personal data to gain social access. Similarly, entertainment apps implement deceptive patterns, such as \textit {grinding}, where users of apps like DramaBox \cite{dramabox} are required to repeatedly complete tasks such as watching ads and inviting others to the app to access locked episodes (Figure \ref{fig:DramaBox Locked Episodes}), often leading to purchases to overcome these barriers \cite{dramabox}. 

While \textit{forced action} is the dominant high-level deceptive pattern among teens, as shown in Figure \ref{fig:all age groups}, they also encounter \textit {social engineering} in entertainment, gaming, shopping, and social media. Social media apps utilized \textit{personalization} patterns, whereas entertainment apps implemented \textit {limited time message} and \textit {countdown timer} patterns, creating a sense of urgency for users to take action. Our results also revealed that teens were less exposed to \textit {sneaking} than adults. We observed \textit {sneaking} implemented only within the shopping app category in contrast to adults, where this deceptive pattern was observed in most app categories except gaming. 

\paragraph{\textbf{Adults}}
Our data revealed that across app categories, apps targeted at adults implement the highest frequency of several high-level deceptive patterns, such as in \textit {forced action} and \textit {social engineering}. \textit {Social engineering} was most frequent in shopping apps, as shown in Figure \ref{fig:Adults HL}. We observed adult shopping apps implementing similar deceptive patterns as teens, such as \textit {limited time message}, \textit {countdown timer}, and \textit {low stock}, all intending to induce purchase urgency. For example, we observed marketing cues in AliExpress \cite{aliexpress} stating ``Only 1 left”, nudging users to quickly interact. 

\begin{figure}[htbp]
         \centering
\includegraphics[width=0.5\textwidth]{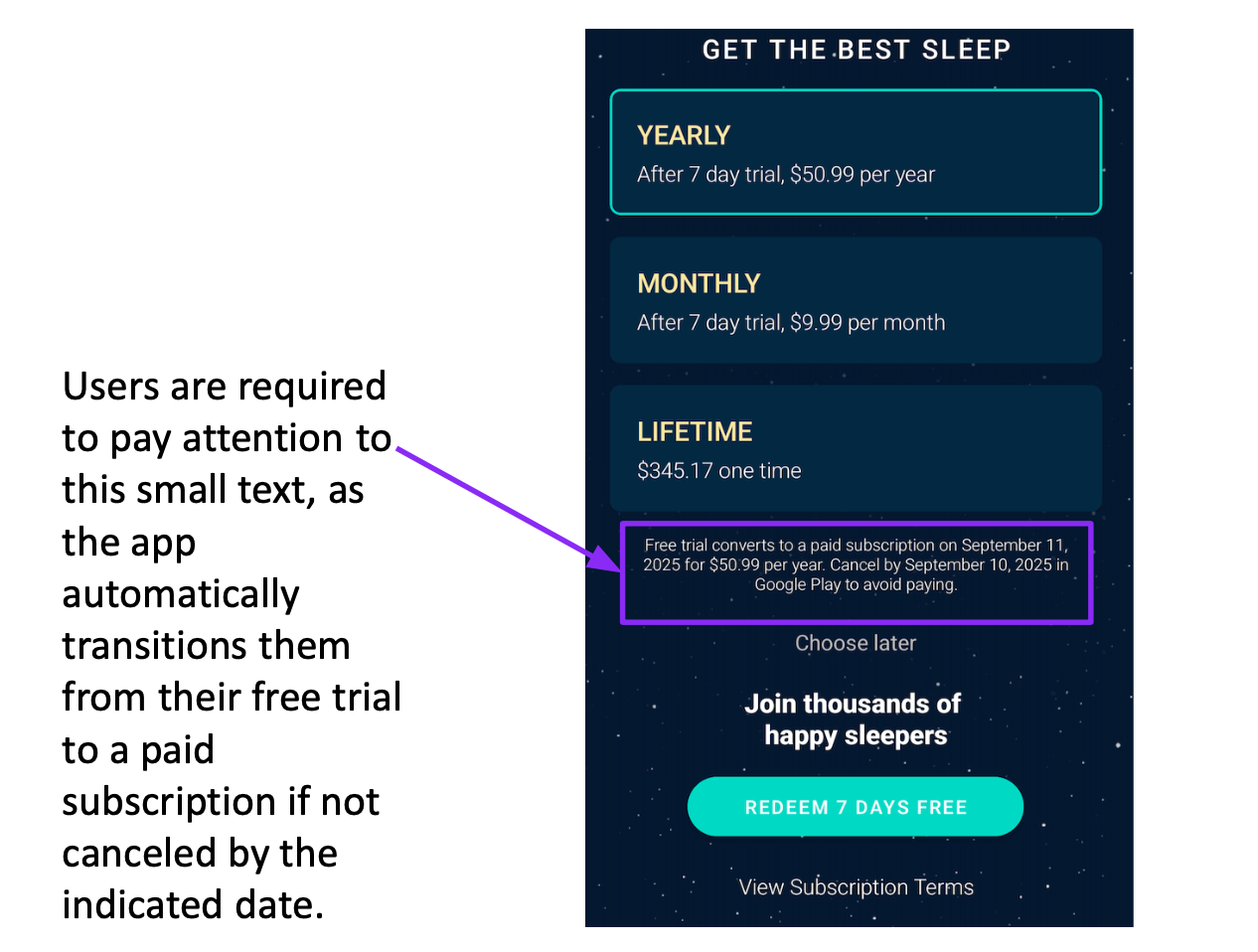}
\caption{\textbf{Slumber\cite{slumber} utilizing a 7-day trial and automatically transitioning users to paid subscription if not cancelled}}
         \label{fig:FC Slumber}
     \end{figure}

Moreover, \textit{forced action} was deployed across all app categories. For example, in the health \& fitness app Slumber \cite{slumber}, the \textit {forced continuity} pattern was implemented through encouraging users to switch to ``premium'' by providing a 7-day free trial, after which a user must make a decision to continue. In a small text after listing the subscription details (see Figure \ref{fig:FC Slumber}), the app highlights the date it will automatically transition to a paid subscription unless a user takes action to cancel. Placing this in small text makes it easy to overlook important information. This also contradicts the user's expectation to get notified before this period, effectively manipulating uninterested users into enrolling in a paid app, which may not be in their favour. Additionally, adults showed \textit {obstruction} in more app categories compared to both teens and older adults (see Figure \ref{fig:Bubble chart}); our data shows that this high-level pattern was observed in gaming apps more frequently compared to \textit {sneaking} patterns. Meanwhile, \textit {interface interference} for health \& fitness apps categorized for adults showed the highest recorded prevalence across all age groups. 

\paragraph{\textbf{Older Adults}} 
Apps targeted towards older adults employed fewer deceptive patterns, compared to teen- and adult-focused apps. Figure \ref{fig:OA HL} shows that older adults are primarily targeted in entertainment and gaming apps. When compared to other age groups, older adults show a higher prevalence of \textit {forced action} in gaming apps encountering designs such as guilt-based prompts through the deceptive pattern \textit {confirmshaming}. Our walkthrough of the Lumosity Brain Training \cite{lumosity} app attempted to persuade us to opt-in to premium options, deceiving users through the use of terms such as ``No, thanks continue to limited". This implies that their experience would be ``limited'' which can emotionally impact users. 

Moreover, entertainment apps for seniors implement sneaking more frequently compared to teen and adult apps. This finding is novel as teens are active users, often spending several hours per day online \cite{rothwell}, yet older adults are more targeted through entertainment apps when compared to these younger groups. For example, our data revealed the use of deceptive patterns such as \textit {information without context} for older adults through the CTV app \cite{ctv}. The app displays a green, colourful box that a user can ``Watch Latest Episode" to gain their attention. However, once selected, a user finds that the episode is locked, requiring either a previously owned subscription or signing up for a subscription.

Similarly, within gaming apps, we observed \textit {forced action} implemented through similar tactics as with teens through \textit {grinding}. During our walkthrough of the gaming app Vita Solitaire for Seniors \cite{VitaSolitaire}, the app offered to gain more rewards (e.g., coins) by watching ads. This burdens a user with tedious tasks requiring time and extra effort, which ultimately discourages them from continuing with the free plan and nudging them to opt for a paid subscription. Overall, deceptive patterns impact each group differently, but many patterns may overlap depending on the app's category and purpose.
    
\subsection{Deceptive Pattern Used Across App Categories}
The app categories we analyzed varied in the number of deceptive patterns they implement. Out of the six app categories, we found that there are specific categories which implement deceptive patterns more than others. Our analysis identified 20 distinctive meso- and low-level deceptive patterns across six entertainment apps, as shown in Figure \ref{fig:all age groups}. In contrast, the music/books category contained only 7, the fewest deceptive patterns analyzed. 

\begin{figure*}[p] 
\centering
\begin{tikzpicture}
\begin{axis}[
cycle list name=color list,
every axis plot/.append style={fill},
    width=15cm,
    y=0.5cm, 
    xbar=0pt,
    xmin=0,
    xmax=1,
    bar width=3pt, 
    enlarge y limits={abs=0.2cm},
    ytick=data,
    y dir=reverse,
    axis line style={draw=none}, 
    tick style={draw=none},
    xlabel={Proportion of deceptive patterns  (Standardized)},
    ylabel={Meso- and low level deceptive patterns },
    ymajorgrids=true,
    grid style={dotted, gray!60},
    symbolic y coords={
        Confirmshaming, Personalization, Limited Time Message, Countdown Timer,
        Activity Messages, Parasocial Pressure, Endorsements and Testimonials,
        Low Stock, High Demand, Auto-Play, Grinding, Pay-to-Play, Social Pyramid,
        Address Book Leeching, Friend Spam, Privacy Zuckering, Forced Registration,
        Forced Continuity, Nagging, Feedforward Ambiguity, Complex Language,
        Wrong Language, Hidden Information, Choice Overload, Trick Questions,
        Positive/Negative Framing, Cuteness, Bad Defaults, Pressured Selling,
        Bundling, Visual Prominence, False Hierarchy, Information without Context,
        Conflicting Information, Reference Pricing, Drip Pricing, Sneak into Basket,
        Disguised Ad, Privacy Maze, Intermediate Currency, Price Comparison Prevention,
        Dead End, Immortal Accounts
    },
    yticklabel style={font=\scriptsize},
    legend style={
        at={(0.5,-0.07)}, 
        anchor=north,
        legend columns=3,
        font=\small
    }
]

\addplot+coordinates {
(0,Confirmshaming) (0.25,Personalization) (0.125,Limited Time Message) (0.125,Countdown Timer) (0.25,Activity Messages) (0,Parasocial Pressure) (0.125,Endorsements and Testimonials) (0.125,Low Stock) (0,High Demand) (0.5,Auto-Play) (0.125,Grinding) (0.125,Pay-to-Play) (0,Social Pyramid) (0.125,Address Book Leeching) (0,Friend Spam) (0,Privacy Zuckering) (0.5,Forced Registration) (0,Forced Continuity) (1,Nagging) (0,Feedforward Ambiguity) (0,Complex Language) (0.125,Wrong Language) (0.25,Hidden Information) (0.125,Choice Overload) (0,Trick Questions) (0.25,Positive/Negative Framing) (0,Cuteness) (0.375,Bad Defaults) (0.125,Pressured Selling) (0,Bundling) (0.75,Visual Prominence) (0,False Hierarchy) (0,Information without Context) (0,Conflicting Information) (0.125,Reference Pricing) (0,Drip Pricing) (0,Sneak into Basket) (0,Disguised Ad) (0.125,Privacy Maze) (0.375,Intermediate Currency) (0,Price Comparison Prevention) (0,Dead End) (0,Immortal Accounts)
};

\addplot+coordinates {
(0,Confirmshaming) (0.352941176,Personalization) (0.058823529,Limited Time Message) (0.117647059,Countdown Timer) (0.176470588,Activity Messages) (0,Parasocial Pressure) (0.117647059,Endorsements and Testimonials) (0.117647059,Low Stock) (0.058823529,High Demand) (0.117647059,Auto-Play) (0.058823529,Grinding) (0,Pay-to-Play) (0.05,Social Pyramid) (0.058823529,Address Book Leeching) (0,Friend Spam) (0,Privacy Zuckering) (0.235294118,Forced Registration) (0.058823529,Forced Continuity) (0.941176471,Nagging) (0,Feedforward Ambiguity) (0,Complex Language) (0,Wrong Language) (0.176470588,Hidden Information) (0.05,Choice Overload) (0,Trick Questions) (0.235294118,Positive/Negative Framing) (0.058823529,Cuteness) (0.235294118,Bad Defaults) (0.058823529,Pressured Selling) (0,Bundling) (0.823529412,Visual Prominence) (0,False Hierarchy) (0.058823529,Information without Context) (0,Conflicting Information) (0.117647059,Reference Pricing) (0.06,Drip Pricing) (0,Sneak into Basket) (0.294117647,Disguised Ad) (0.235294118,Privacy Maze) (0.058823529,Intermediate Currency) (0,Price Comparison Prevention) (0.117647059,Dead End) (0,Immortal Accounts)
};

\addplot+ coordinates {
(0.2,Confirmshaming) (0.2,Personalization) (0,Limited Time Message) (0,Countdown Timer) (0,Activity Messages) (0,Parasocial Pressure) (0,Endorsements and Testimonials) (0,Low Stock) (0,High Demand) (0.2,Auto-Play)(0.2,Grinding) (0,Pay-to-Play) (0,Social Pyramid) (0,Address Book Leeching) (0,Friend Spam) (0,Privacy Zuckering) (0.2,Forced Registration) (0,Forced Continuity) (0.8,Nagging) (0.2,Feedforward Ambiguity) (0,Complex Language) (0,Wrong Language) (0,Hidden Information) (0.2,Choice Overload) (0,Trick Questions) (0.4,Positive/Negative Framing) (0,Cuteness) (0.2,Bad Defaults) (0,Pressured Selling) (0,Bundling) (0.4,Visual Prominence) (0,False Hierarchy) (0.2,Information without Context) (0,Conflicting Information) (0,Reference Pricing) (0,Drip Pricing) (0,Sneak into Basket) (0.2,Disguised Ad) (0.2,Privacy Maze) (0.2,Intermediate Currency) (0,Price Comparison Prevention) (0,Dead End) (0,Immortal Accounts)

};

\legend{Teens, Adults, Older Adults}

\end{axis}
\end{tikzpicture}
\caption{\textbf {Meso- and low-level deceptive patterns occurrence across all age groups}}
\label{fig:clustered graph}
\end{figure*}
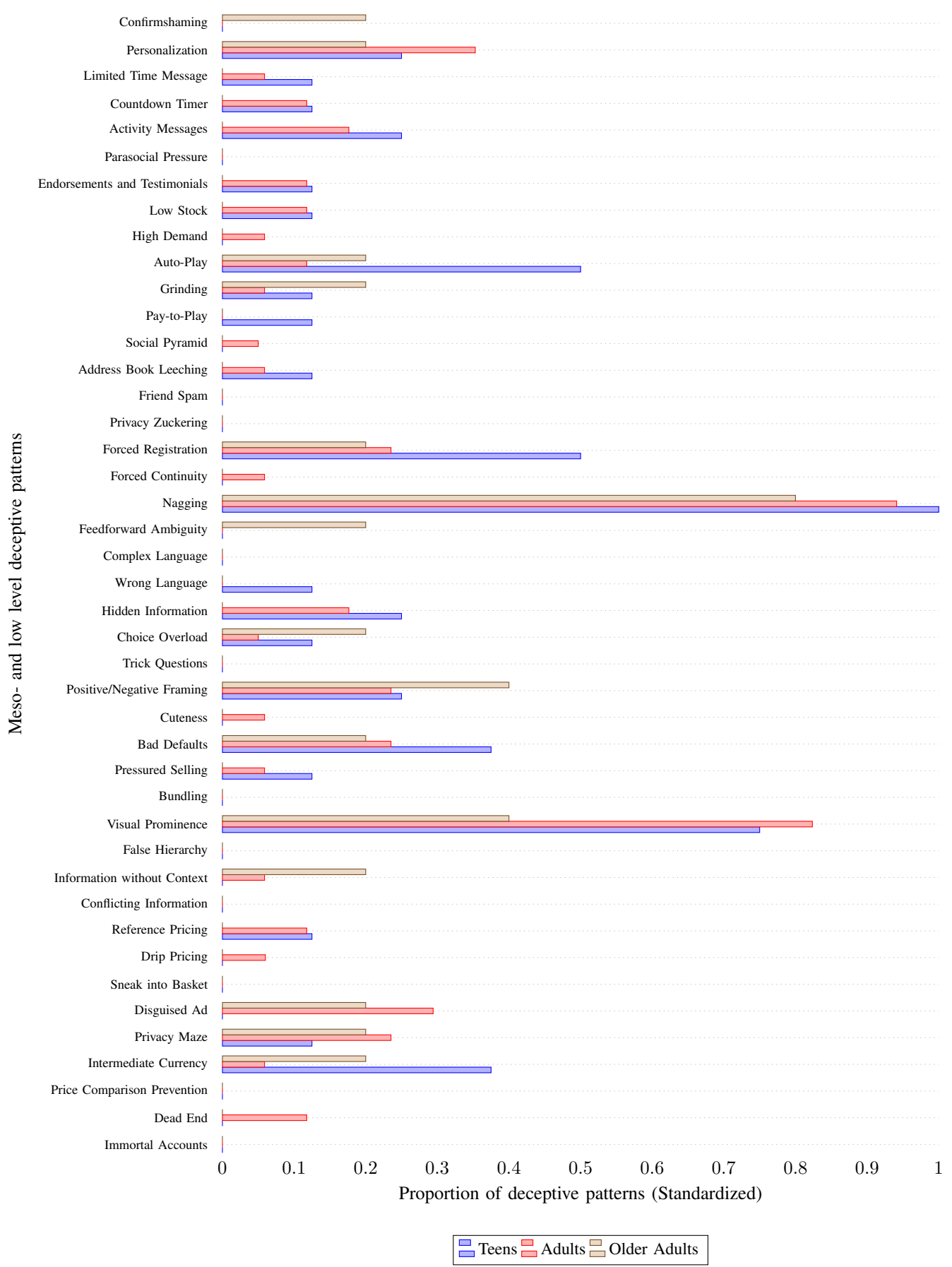

\paragraph{\textbf{Entertainment apps}} 
During our analysis, aside from the commonly observed deceptive patterns (Figure \ref{fig:clustered graph}) (\textit{nagging}, \textit{visual prominence}, \textit{forced registration}, \textit{positive or negative framing}, \textit{personalization}, and \textit{bad defaults}), we observed the \textit{auto-play} pattern implemented more frequently in entertainment apps. This deceptive pattern was present in many of the entertainment apps, for example, DramaBox, YouTube Kids, Netflix, and CTV. For this particular deceptive pattern, the tactic was employed in the same way across apps: the apps automatically transitioned us to the next content (e.g., video) without our explicit confirmation. This acts as a behavioural nudge for users to spend more time on apps than intended. It can also expose users to content that may be harmful or inappropriate for their age. 

\paragraph{\textbf{Music/Books apps}}
The Spotify app \cite{spotify}, for example, incorporated 4 out of 7 meso- and low-level deceptive patterns identified from our walkthrough, including \textit{dead end, visual prominence, auto-play, and bad defaults}. We found that this app category implemented the fewest number of deceptive patterns. Moreover, we came across many ads while exploring Youtube Music \cite{youtube_music}. These ads were embedded as part of the selected video and automatically played between music videos. Burdening users with repetitive interruptions could lead them to succumb to the temptation of upgrading to the premium version to stop ads. Despite including some of the most common deceptive patterns observed, this app category was not found to implement the \textit{social engineering} pattern. Figure \ref{fig:all age groups} shows a higher occurrence of \textit{social engineering} in other app categories such as shopping and entertainment in contrast to music/books apps. This suggests that while some deceptive patterns are commonly employed, the type of tactic is largely dependent on the targeted age group and app category.

\begin{tcolorbox}[colback=gray!10, colframe=gray!50]
Across all age groups and app categories, \textit{nagging} and \textit{visual prominence} are the most prevalent meso- and low-level deceptive patterns. Teens encounter more frequent \textit{forced action} patterns in social media and entertainment apps. Adults demonstrate the highest exposure to several high-level deceptive patterns, such as \textit{obstruction}, while apps targeted at older adults generally deploy fewer deceptive patterns.  

\end{tcolorbox}

\begin{tcolorbox}[colback=gray!10, colframe=gray!50]
The entertainment app category was found to implement the most deceptive patterns overall. By comparison, the music/books category demonstrated the fewest deceptive patterns across all age groups and was not observed to utilize \textit{social engineering} patterns. 

\end{tcolorbox}

\section{Discussion}

Through our heuristic-based cognitive walkthrough, we observed that deceptive patterns are not deployed the same across mobile apps, but rather appear strategically implemented depending on the app's intended age group, purpose, and monetization goals. Although \textit{nagging, visual prominence, personalization, and forced registration} were the most frequent deceptive patterns observed in our study, the way they were executed as designs varied based on the target age groups and app categories. These results suggest that deceptive patterns are not randomly distributed but are carefully constructed to influence user decision-making by nudging them to take a specific action. 

\subsection{{Influence of Deceptive Patterns on App Categories}} 
The entertainment app category contained the most deceptive patterns. We argue that marketing strategies like focusing on capturing attention have become deeply inherent components of an app's structure, rather than just add-on elements. Entertainment apps, often adopting strategies from mobile gaming, heavily focus on user retention and increasing engagement through the implementation of freemium models, and slowly encourage users to opt in for subscriptions to avoid barriers like ads \cite{alha2014}. Entertainment apps benefit from such designs as users may engage with offers or subscribe to the provided services, increasing opportunities for monetization and tracking of user behaviours to extend interaction. 

This trend aligns with Rappold \cite{rappold2025monetization_online}, who highlights that monetization is one of the main factors aiding in mobile app sustainability. However, our analysis of this business model suggests that financial pressure often leads to monetization at the cost of user experience. Supporting this angle, Kitkowska \cite{Kitkowska2023} highlighted that effectively implemented patterns influence users’ decision-making abilities through UI designs and visual elements such as colour and layout. We observed this frequently in entertainment apps, for example, through \textit{nagging} and \textit{visual prominence}. By contrast, the lack of deceptive patterns in music/books apps suggests that business models that focus on long-term usage by employing ad-supported freemium models \cite{zyneto} may have less incentive to implement manipulative strategies. Although monetization remains a common goal among apps, the music/books category may often focus on building long-term habits and user trust, suggesting a possibly safer environment for users, which researchers can further explore.

\subsection{{Vulnerability Across Age Groups}} 
One of our main findings gained from our walkthrough is how deceptive patterns impact age-specific weaknesses.

\paragraph{{\textbf{Teens and Cognitive Burden}}} 
Teen-targeted apps frequently incorporate \textit{interface interference} patterns, employing \textit{visual prominence, bad default, and positive or negative framing} to steer user behaviour. While these patterns are deployed to persuade younger users towards paid tiers or sharing data, their true influence depends on how they exploit distinctive adolescent cognitive and social vulnerabilities. According to Orben and Blakemore \cite{OrbenandBlakemore}, teens place more emphasis on peer approval, fearing rejection and digital exclusion impacting their mood compared to other age groups.    

Deceptive patterns often exploit this social anxiety by persuading users to upgrade to paid plans and make purchases to enhance their experience and gain social benefits. Prior work by Kusmawan et al. \cite{Kusmawanetal2026} highlights how virtual currencies drive continuous payment cycles among young users, encouraging them to purchase items such as clothes and accessories to gain social status. This implies that such designs can contribute to social pressure to obtain special status among their communities. Similarly, we observed that apps like DramaBox \cite{dramabox} are deploying deceptive patterns such as \textit{grinding} to directly influence young users' desire for social status. This finding proves that apps are strategically exploiting teens' developing impulse control rather than deploying these patterns randomly. 

In parallel, designers manipulate a distinct adolescent vulnerability by not accounting for their cognitive control development. This, for instance, makes teens particularly vulnerable to visual deceptive patterns such as invisible, 'X' close buttons or disguised ads integrated in videos. Such designs do not directly guide social status but rather deceive teens developing impulse control, nudging them to unintended interactions within digital environments, overall leading to cognitive burdens and pressure \cite{Chamorro2024}.

\paragraph{\textbf{Adults and Revenue-Driven Tactics}} 
Adults, in contrast to teens and older adults,  
are a primary target for \textit{social engineering} and \textit{forced action} strategies. This age group is considered a primary revenue-generating group—accounting for the highest subscription uptake for streaming services \cite{subscription_gap}. Designers employ various patterns impacting this age group, such as \textit{nagging, high demand, low stock, and forced registration} to manipulate users by causing a sense of urgency to act. This indicates that while teens experience social Fear of Missing Out (FOMO), adult-oriented apps are developed to exploit financial FOMO \cite{hodkinson2019}, leading consumers to contribute financially by opting into deals. This is supported by our findings as adults are exploited through the deceptive pattern \textit{forced continuity} in the Slumber app \cite{slumber}. 

\paragraph{\textbf{Older Adults and Deceptive Reassurance}} 
Older adult apps in our study included fewer deceptive patterns overall. However, our research implies that older adults are still vulnerable when utilizing mainstream apps—such as those related to shopping or entertainment—which exposes them to deceptive patterns that often do not meet their cognitive needs. In contrast to younger users who may navigate complex platforms, older adults encounter a particular risk as mainstream apps exploit factors like cognitive processing limitations. Prior research revealed that there is a strong effect of deceptive patterns on older adults’ ability to understand they are being deceived, and that they are even less aware of how their behaviours are being influenced \cite{Blanchy2021}. Our findings demonstrate this through the CTV app \cite{ctv} where older adults are misled coming across the ``Watch Latest Episode" prompt. This suggests a vulnerability gap among this age group, where the apps they most use (mainstream apps) fail to deploy age-inclusive design standards.

\subsection{{Reflection for Future Research}} 
While conducting this heuristic-based cognitive walkthrough, we faced some challenges which we detail here to benefit future researchers. First, we faced difficulty in identifying the target age group for multiple apps; despite some being obvious, the majority required external research into apps' marketing materials and Google Play's \cite{google_play} target audience indicator details. To address this challenge, we recommend that future research go beyond assessing apps for intended age groups by shifting the methodological focus to implementing personas. Deploying specific personas targeting each age group allows researchers to study apps focusing on user vulnerabilities, disregarding the app store's age recommendation classification mechanism. This shifts the focus onto behavioural effects, such as the cognitive gap for older adults or teens' vulnerability to social pressure, leading to impulsive decisions to gain social status.

Additionally, in some instances, we had challenges verifying the designer's intent. For certain deceptive patterns identified during our walkthrough, it was difficult to confidently assert whether we were encountering an explicit attempt to deceive users or if we had just identified poor app design elements. For example, with the AutoTrader app \cite{autotrader} during our walkthrough, we attempted to review their expert advice service on how to sell as a private owner; however, we continuously found an error 404 stating this page may not exist. This made it challenging to identify whether the designer's intent was to prevent us from accessing this information or simply a technical glitch. To mitigate this uncertainty, we heavily relied on the framework provided by Gray et al. \cite{gray2024ontology}. This ensured that when coming across a potential deceptive pattern, we used the categorized patterns and their definitions to ensure alignment with their strategic purposes. Moreover, we recommend future researchers working in this field to focus more on meso-level deceptive patterns  \cite{gray2024ontology}. While low-level deceptive patterns, such as text size, make it easier to spot certain manipulative attempts, our results show that the frequently deployed patterns in our study fell under the meso-level. Focusing on studying patterns under this category provides deeper insights to better understand the intent of designers, since these frameworks demonstrate the systematic nature of deception. By prioritizing meso-level patterns, researchers will have the ability to understand long-term deceptive influences while app UI design trends may evolve.

\section{Conclusion and Future Work}
Overall, our research shows that deceptive patterns are strategically implemented to target different age groups (teens, adults, and older adults), exploiting age-specific cognitive and behavioural vulnerabilities across different app categories. Our heuristic-based cognitive walkthrough identified that adult-targeted apps deploy the most deceptive patterns, with entertainment apps showing the most implementation across all age groups. Additionally, we found that \textit{forced action} and \textit{interface interference} are frequently used to deceive various users, regardless of app categories and age groups. Our future research will examine how these findings are perceived by gathering actual user insights to better understand the gaps between age groups and analyze their behaviour when encountering deception. These insights can support ongoing efforts to design tools, methods, and age-inclusive design frameworks that minimize the impact of deceptive designs across various app categories. 

\bibliographystyle{IEEEtran}
\bibliography{ref}

\appendix

\definecolor{heatRed}{HTML}{F27272}
\definecolor{heatOrangeRed}{HTML}{F7926C}
\definecolor{heatOrange}{HTML}{FFB366}
\definecolor{heatTan}{HTML}{FFD685}
\definecolor{heatYellow}{HTML}{FFEB9C}

\begin{figure}[htbp] 
\centering
\footnotesize 

\setlength{\tabcolsep}{2.5pt} 

\renewcommand{\arraystretch}{1.15}

\begin{NiceTabular*}{\columnwidth}{clcccccc}[hvlines-except-borders, cell-space-limits = 2pt]
\RowStyle{\bfseries}
HL Cat. & Meso/Low Pattern & Ent. & Gam. & H\&F & M/B & Sho. & SM. \\

\Block{9-1}{\rotatebox{90}{\textbf{Social Eng.}}} 
& Confirmshaming &   & \cellcolor{heatYellow} 0.2 &   &   &   &   \\
& Personalization &   &  & \cellcolor{heatOrange} 0.6 &  & \cellcolor{heatOrangeRed} 0.8 & \cellcolor{heatTan} 0.4 \\
& Limited Time Msg. & \cellcolor{heatYellow} 0.2 &   &   &   & \cellcolor{heatYellow} 0.2 &   \\
& Countdown Timer & \cellcolor{heatYellow} 0.2 & \cellcolor{heatYellow} 0.2 &   &   & \cellcolor{heatYellow} 0.2 &   \\
& Activity Messages & \cellcolor{heatYellow} 0.2 & \cellcolor{heatYellow} 0.2 &   &   & \cellcolor{heatOrange} 0.6 &  \\
& Parasocial Pressure &   &   &  &  &   &   \\
& Endorsements &   &   &   &   & \cellcolor{heatOrange} 0.6 &   \\
& Low Stock & \cellcolor{heatYellow} 0.2 &  &  &  & \cellcolor{heatTan} 0.4 &   \\
& High Demand &   &   &   &   & \cellcolor{heatYellow} 0.2 &   \\
\hline

\Block{10-1}{\rotatebox{90}{\textbf{Forced Action}}} 
& Auto-Play & \cellcolor{heatOrangeRed} 0.8 &   &   & \cellcolor{heatTan} 0.4 &   & \cellcolor{heatYellow} 0.2 \\
& Grinding & \cellcolor{heatYellow} 0.2 & \cellcolor{heatTan} 0.4 &   &   &   &   \\
& Pay-to-Play & \cellcolor{heatYellow} 0.2 &   &  &   &   &   \\
& Social Pyramid &   &   & \cellcolor{heatYellow} 0.2 &   &   &   \\
& Address Book Leech. &   &   &   &   &   & \cellcolor{heatTan} 0.4 \\
& Friend Spam &   &   &   &   &   &   \\
& Privacy Zuckering &  &   &   &   &   &   \\
& Forced Reg. & \cellcolor{heatOrange} 0.6 &   &   &   & \cellcolor{heatOrange} 0.6 & \cellcolor{heatOrange} 0.6 \\
& Forced Continuity &   &   & \cellcolor{heatYellow} 0.2 &   &   &   \\
& Nagging & \cellcolor{heatRed} 1.0 & \cellcolor{heatRed} 1.0 & \cellcolor{heatRed} 1.0 & \cellcolor{heatOrangeRed} 0.8 & \cellcolor{heatRed} 1.0 & \cellcolor{heatRed} 1.0 \\
\hline

\Block{13-1}{\rotatebox{90}{\textbf{Interface Interf.}}}
& Feedforward Ambig. &   & \cellcolor{heatYellow} 0.2 &   &   &   &  \\
& Complex Language &   &   &   &   &   &   \\
& Wrong Language & \cellcolor{heatYellow} 0.2 &   &   &   &   &   \\
& Hidden Information & \cellcolor{heatYellow} 0.2 & \cellcolor{heatYellow} 0.2 & \cellcolor{heatYellow} 0.2 & \cellcolor{heatYellow} 0.2 &  & \cellcolor{heatYellow} 0.2 \\
& Choice Overload & \cellcolor{heatTan} 0.4 &   &   &   &  &   \\
& Trick Questions &   &   &   &  &   &   \\
& Pos./Neg. Framing & \cellcolor{heatTan} 0.4 & \cellcolor{heatOrange} 0.6 & \cellcolor{heatOrange} 0.6 &   &   & \cellcolor{heatYellow} 0.2 \\
& Cuteness &  & \cellcolor{heatYellow} 0.2 &  &  &   &   \\
& Bad Defaults & \cellcolor{heatOrange} 0.6 & \cellcolor{heatYellow} 0.2 &   & \cellcolor{heatTan} 0.4 &   & \cellcolor{heatOrange} 0.6 \\
& Pressured Selling & \cellcolor{heatYellow} 0.2 &  & \cellcolor{heatYellow} 0.2 &   &   &  \\
& Bundling &   &   &   &   &   &   \\
& Visual Prominence & \cellcolor{heatOrangeRed} 0.8 & \cellcolor{heatTan} 0.4 & \cellcolor{heatOrangeRed} 0.8 & \cellcolor{heatOrangeRed} 0.8 & \cellcolor{heatOrangeRed} 0.8 & \cellcolor{heatOrangeRed} 0.8 \\
& False Hierarchy &   &   &   &   &   &  \\
\hline

\Block{6-1}{\rotatebox{90}{\textbf{Sneaking}}}
& Info. w/o Context & \cellcolor{heatYellow} 0.2 &   & \cellcolor{heatYellow} 0.2 &   &   &   \\
& Conflicting Info. &   &   &   &   &   &   \\
& Reference Pricing &   &   &   &   & \cellcolor{heatOrange} 0.6 &  \\
& Drip/Hidden/Part. &   &   &   &   &   & \cellcolor{heatYellow} 0.2 \\
& Sneak into Basket &   &  &   &   &   &   \\
& Disguised Ad & \cellcolor{heatTan} 0.4 & \cellcolor{heatYellow} 0.2 & \cellcolor{heatYellow} 0.2 & \cellcolor{heatYellow} 0.2 & \cellcolor{heatTan} 0.4 &   \\
\hline

\Block{5-1}{\rotatebox{90}{\textbf{Obstruction}}}
& Privacy Maze & \cellcolor{heatOrange} 0.6 & \cellcolor{heatYellow} 0.2 & \cellcolor{heatYellow} 0.2 &  & \cellcolor{heatYellow} 0.2 &   \\
& Intermed. Currency &   & \cellcolor{heatOrange} 0.6 &   &  &  & \cellcolor{heatYellow} 0.2 \\
& Price Comp. Prev. &   &   &   &   &   &   \\
& Dead End &   &   &   & \cellcolor{heatYellow} 0.2 & \cellcolor{heatYellow} 0.2 &   \\
& Immortal Accounts &   &   &   &   &   &   \\

\end{NiceTabular*}

    \caption{\textbf{Full occurrence heatmap of 43 deceptive patterns}}
    \label{fig:meso-low-dp} 
\end{figure}

\clearpage

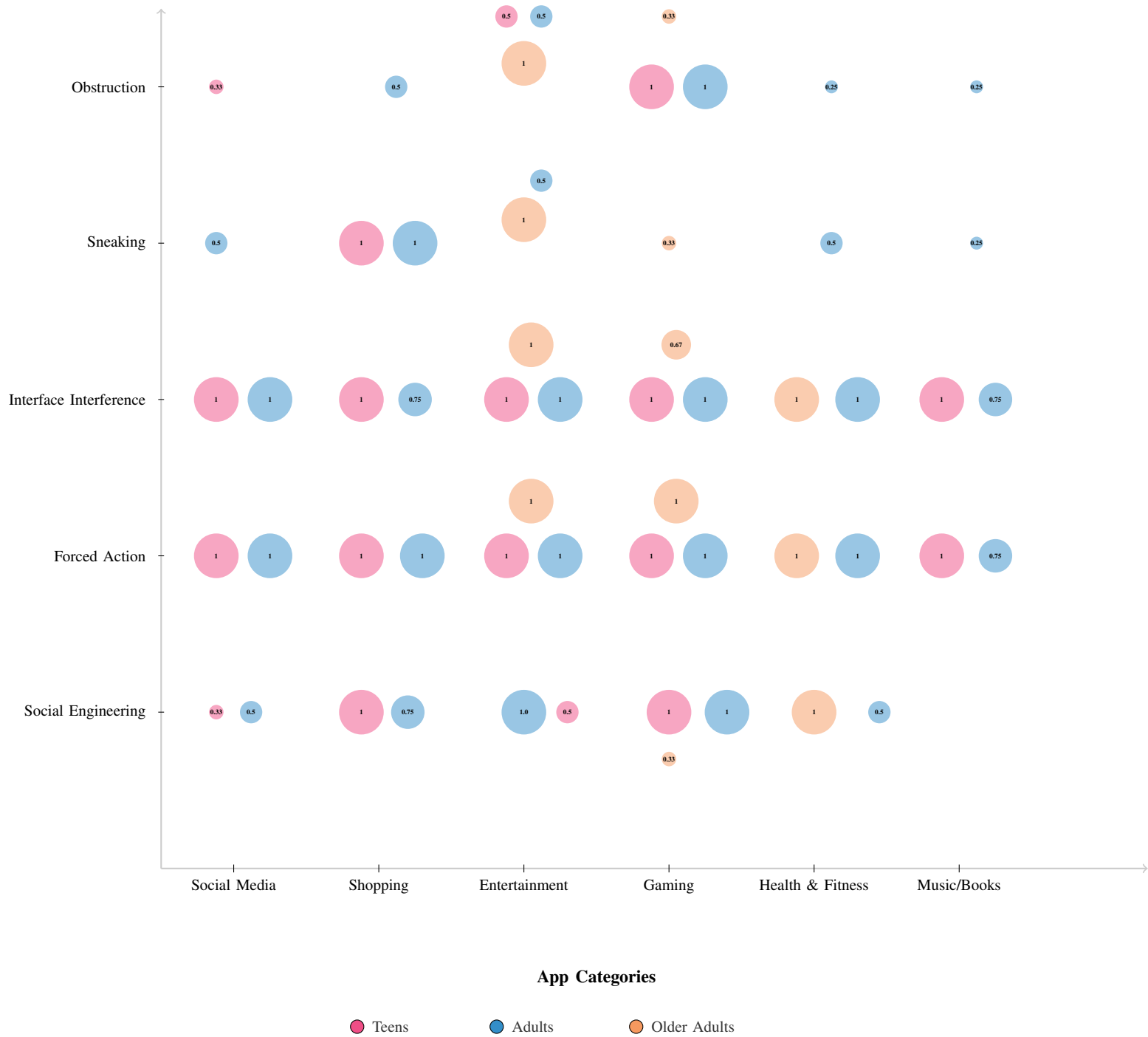
\begin{figure*}[ht]
    \centering
    \hspace*{-1.8cm} 
    \begin{tikzpicture}[x=2.6cm, y=2.8cm]
    
    \draw[thick, ->, gray!40] (-0.5,0) -- (6.3,0);
    \draw[thick, ->, gray!40] (-0.5,0) -- (-0.5,5.5);

 \node[font=\small\bfseries] at (2.5, -0.7) {App Categories};
  \node[font=\small\bfseries, rotate=90] at (-1.9, 2.75) {High-Level Dark Patterns};

\foreach \y/\label in {1/Social Engineering, 2/Forced Action, 3/Interface Interference, 4/Sneaking, 5/Obstruction}
    \draw (-0.48, \y) -- (-0.52, \y) node[left, font=\footnotesize, xshift=-0.1cm] {\label};

     \foreach \x/\label in {0/{Social Media}, 1/{Shopping}, 2/{Entertainment}, 3/{Gaming}, 4/{Health \& Fitness}, 5/{Music/Books}}
        \draw (\x, 2pt) -- (\x, -2pt) node[below, font=\footnotesize, align=center] {\label};

     \tikzset{
         teen/.style={circle, fill=WildStrawberry, fill opacity=0.4, text opacity=1, text=black, font=\fontsize{3.5pt}{3.5pt}\selectfont\bfseries, inner sep=0pt},
        adult/.style={circle, fill=RoyalBlue, fill opacity=0.4, text opacity=1, text=black, font=\fontsize{3.5pt}{3.5pt}\selectfont\bfseries, inner sep=0pt},
        older/.style={circle, fill=Orange, fill opacity=0.4, text opacity=1, text=black, font=\fontsize{3.5pt}{3.5pt}\selectfont\bfseries, inner sep=0pt}
    }

    Y=1: SOCIAL ENGINEERING
    
    \node[teen,  minimum size=0.26cm] at (-0.12, 1.0) {0.33}; 
    \node[adult, minimum size=0.4cm]  at (0.12, 1.0) {0.5};   
    \node[teen,  minimum size=0.8cm]  at (0.88, 1.0) {1};     
     \node[adult, minimum size=0.6cm]  at (1.20, 1.0) {0.75};  
     \node[teen,  minimum size=0.4cm]  at (2.30, 1.0) {0.5};   
    \node[adult, minimum size=0.8cm]  at (2.0, 1.0) {1.0};    
    \node[teen,  minimum size=0.8cm]  at (3.0, 1.0) {1};     
    \node[adult, minimum size=0.8cm]  at (3.4, 1.0) {1};     
    \node[older, minimum size=0.26cm] at (3.0, 0.70) {0.33};  
    \node[adult, minimum size=0.4cm]  at (4.45, 1.0) {0.5};   
    \node[older, minimum size=0.8cm]  at (4.0, 1.0) {1};     

    \node[teen,  minimum size=0.8cm]  at (-0.12, 2.0) {1};    
    \node[adult, minimum size=0.8cm]  at (0.25, 2.0) {1};     
    \node[teen,  minimum size=0.8cm]  at (0.88, 2.0) {1};     
    \node[adult, minimum size=0.8cm]  at (1.3, 2.0) {1};     
    \node[teen,  minimum size=0.8cm]  at (1.88, 2.0) {1};     
    \node[adult, minimum size=0.8cm]  at (2.25, 2.0) {1};     
    \node[older, minimum size=0.8cm]  at (2.05, 2.35) {1};     
    \node[teen,  minimum size=0.8cm]  at (2.88, 2.0) {1};     
    \node[adult, minimum size=0.8cm]  at (3.25, 2.0) {1};     
    \node[older, minimum size=0.8cm]  at (3.05, 2.35) {1};     
    \node[adult, minimum size=0.8cm]  at (4.30, 2.0) {1};     
    \node[older, minimum size=0.8cm]  at (3.88, 2.0) {1};     
    \node[teen,  minimum size=0.8cm]  at (4.88, 2.0) {1};     
    \node[adult, minimum size=0.6cm]  at (5.25, 2.0) {0.75};  

    \node[teen,  minimum size=0.8cm]  at (-0.12, 3.0) {1};    
    \node[adult, minimum size=0.8cm]  at (0.25, 3.0) {1};     
    \node[teen,  minimum size=0.8cm]  at (0.88, 3.0) {1};     
    \node[adult, minimum size=0.6cm]  at (1.25, 3.0) {0.75};  
    \node[teen,  minimum size=0.8cm]  at (1.88, 3.0) {1};     
    \node[adult, minimum size=0.8cm]  at (2.25, 3.0) {1};     
    \node[older, minimum size=0.8cm]  at (2.05, 3.35) {1};     
    \node[teen,  minimum size=0.8cm]  at (2.88, 3.0) {1};     
    \node[adult, minimum size=0.8cm]  at (3.25, 3.0) {1};     
    \node[older, minimum size=0.53cm] at (3.05, 3.35) {0.67};  
    \node[adult, minimum size=0.8cm]  at (4.30, 3.0) {1};     
    \node[older, minimum size=0.8cm]  at (3.88, 3.0) {1};     
    \node[teen,  minimum size=0.8cm]  at (4.88, 3.0) {1};     
    \node[adult, minimum size=0.6cm]  at (5.25, 3.0) {0.75};  

    \node[adult, minimum size=0.4cm]  at (-0.12, 4.0) {0.5};   
    \node[teen,  minimum size=0.8cm]  at (0.88, 4.0) {1};     
    \node[adult, minimum size=0.8cm]  at (1.25, 4.0) {1};     
    \node[adult, minimum size=0.4cm]  at (2.12, 4.40) {0.5};  
    \node[older, minimum size=0.8cm]  at (2.0, 4.15) {1};     
    \node[older, minimum size=0.26cm] at (3.0, 4.0) {0.33};   
    \node[adult, minimum size=0.4cm]  at (4.12, 4.0) {0.5};   
    \node[adult, minimum size=0.2cm]  at (5.12, 4.0) {0.25};  

    \node[teen,  minimum size=0.26cm] at (-0.12, 5.0) {0.33}; 
    \node[adult, minimum size=0.4cm]  at (1.12, 5.0) {0.5};   
    \node[teen,  minimum size=0.4cm]  at (1.88, 5.45) {0.5};  
    \node[adult, minimum size=0.4cm]  at (2.12, 5.45) {0.5};  
    \node[older, minimum size=0.8cm]  at (2.0, 5.15) {1};     
    \node[teen,  minimum size=0.8cm]  at (2.88, 5.0) {1};     
    \node[adult, minimum size=0.8cm]  at (3.25, 5.0) {1};     
    \node[older, minimum size=0.26cm] at (3.0, 5.45) {0.33};  
    \node[adult, minimum size=0.2cm]  at (4.12, 5.0) {0.25};  
    \node[adult, minimum size=0.2cm]  at (5.12, 5.0) {0.25};  

    \end{tikzpicture}

    \vspace{4mm}
    \centering
    \begin{tikzpicture}
        \draw[fill=WildStrawberry, fill opacity=0.8] (0,0) circle (0.12cm) node[right=0.15cm, font=\footnotesize] {Teens};
        \draw[fill=RoyalBlue, fill opacity=0.8] (2.5,0) circle (0.12cm) node[right=0.15cm, font=\footnotesize] {Adults};
        \draw[fill=Orange, fill opacity=0.8] (5,0) circle (0.12cm) node[right=0.15cm, font=\footnotesize] {Older Adults};
    \end{tikzpicture}

\caption{\textbf{Comparative analysis of high-level dark patterns across all age groups and app categories}}
    \label{fig:Bubble chart}
\end{figure*}

\end{document}